\documentclass
[twocolumn,showpacs,preprintnumbers,amsmath,amssymb,prc]{revtex4}
\usepackage{graphicx}
\usepackage{dcolumn}
\usepackage{bm}

\begin{document}

\title{Effect of particle fluctuation on isoscaling and isobaric yield ratio of nuclear multifragmentation}

\author{ Swagata Mallik and Gargi Chaudhuri}

\affiliation{Theoretical Physics Division, Variable Energy Cyclotron Centre,
1/AF Bidhan Nagar, Kolkata700064,India}

\begin{abstract}
Isoscaling and isobaric yield ratio parameters are compared from canonical and grand canonical
ensembles when applied to multifragmentation of finite nuclei. Source dependence of isoscaling
 parameters \& source and isospin dependence of isobaric yield ratio parameters are examined in
the framework of the canonical and the grand canonical models. It is found that as the nucleus fragments
more, results from both the ensembles converge and observables calculated from the canonical ensemble coincide
more with those obtained from the formulae derived using the grand canonical ensemble.
\end{abstract}
\pacs{25.70.Mn, 25.70.Pq}
\maketitle

\section{Introduction}
The study of the nuclear equation of state is an important area of research in intermediate energy heavy ion reactions \cite{Bao-an-li1, Bao-an-li2} and results from nuclear multifragmentation reactions are extensively used for such study. The statistical models are extremely powerful and widely used tools for study of the multifragmentation reactions. In models of statistical disassembly of a nuclear system formed by the collision of two heavy ions at intermediate energy
one assumes that the hot and compressed nuclear system expands and subsequently fragments into composites
of different masses depending on the initial conditions. The fragmentation of the nucleus into available
 channels (depends on phase space) can be solved in different statistical ensembles (microcanonical,
 canonical and grand canonical). For finite nuclei, in general the results for different observables differ
 in different ensembles and they are found to converge under certain conditions \cite{Mallik4}.\\
\indent
Isoscaling \cite{Tsang1,Botvina1,Chaudhuri4,Colonna,Raduta,Mallik5} and isobaric yield
ratio \cite{Mallik5,Huang,Tsang2,Ma1} are two well known methods which are
used to study the nuclear EOS and to extract liquid drop model parameters
(symmetry energy coefficient for example) from multifragmentation reactions.
  The isoscaling parameters are related to the difference between the chemical potentials
 of the two fragmenting isotopes and hence provide insight into their symmetry energy.
 The isobaric yield ratio method is used to extract the liquid drop model parameters
 and also the difference in chemical potential between neutrons and protons. The formulae
for both these methods are derived using the framework of the grand canonical model.
 Hence results from this ensemble agree exactly with those from these equations. The canonical model
 is better suited compared to grand canonical model for describing intermediate energy nuclear reactions
 where baryon and charge numbers are conserved. On the contrary, results from the canonical ensemble differ from the formulae based on grand canonical ensemble.
The nature of this deviation depends on the size of the fragmenting systems as well as on the
 neutron to proton ratio (asymmetry) of them.
 It is seen that results from both the ensembles differ when
 the source size is small and asymmetry is large and they come closer when one increases
 the source size or decrease its asymmetry.
The results of the isoscaling as well as the isobaric yield ratio equations depend on the statistical ensemble used for the calculation. Since the equations for extracting the relevant parameters are derived from the grand canonical framework, hence the results from the canonical model should be carefully analyzed for extraction of different parameters from these methods.\\
\indent
The letter is structured as follows. In section II, we give a brief introduction to the canonical
 and the grand canonical models where as section III contains the theoretical framework of
 isoscaling and isobaric yield ratio methods from grand canonical ensemble.
 The results  will  be presented in Section IV. Finally we shall summarize and conclude in section V.\\
\section{The Canonical and the Grand Canonical Model}
In this section we describe briefly
the canonical and the grand canonical models of nuclear multifragmentation.
The basic output from canonical \cite{Das1} or grand canonical model \cite{Chaudhuri1} is multiplicity
of the fragments. After calculating the multiplicities, isoscaling and isobaric yield ratio parameters can be obtained. By multiplicity we mean that the average
number of fragments produced for each proton number $Z$ and neutron
number $N$. Assuming that the system with $A_{0}$ nucleons and $Z_{0}$
protons at temperature $T$, has expanded to a volume higher than
normal nuclear volume, the partitioning into different composites
can be calculated according to the rules of equilibrium statistical
mechanics.\\
\indent
In a canonical model \cite{Das1}, the partitioning is done such that all partitions have the correct $A_{0},Z_{0}$ (equivalently $N_{0},Z_{0}$). The canonical partition function is given by
\begin{eqnarray}
Q_{N_{0},Z_{0}} & = & \sum\prod\frac{\omega_{N,Z}^{n_{N,Z}}}{n_{N,Z}!}
\end{eqnarray}
where the sum is over all possible channels of break-up (the number of such channels is enormous) satisfying $N_{0}=\sum N\times n_{N,Z}$ and $Z_{0}=\sum Z\times n_{N,Z}$; $\omega_{N,Z}$ is the partition function of the composite with $N$ neutrons \& $Z$ protons and $n_{NZ}$ is its multiplicity. The partition function $Q_{N_{0},Z_{0}}$ is calculated using a recursion relation \cite{Das1}. From Eq. (1), the average number of composites is given by \cite{Das1}
\begin{eqnarray}
\langle n_{N,Z}\rangle_{c} & = & \omega_{N,Z}\frac{Q_{N_{0}-N,Z_{0}-Z}}{Q_{N_{0},Z_{0}}}
\end{eqnarray}
It is necessary to specify which nuclei are included in computing $Q_{N_{0},Z_{0}}$. For $N,Z$ we include a ridge along the line of stability. The liquid-drop formula gives neutron and proton drip lines and the results shown here include all nuclei within the boundaries.\\
\indent
In the grand canonical model \cite{Chaudhuri1}, if the neutron chemical potential is $\mu_{n}$ and the proton chemical potential is $\mu_{p}$, then statistical equilibrium implies \cite{Reif} that the chemical potential of a composite with $N$ neutrons and $Z$ protons is $\mu_{n}N+\mu_{p}Z$.
The average number of composites with $N$ neutrons and $Z$ protons is given by \cite{Chaudhuri1}
\begin{eqnarray}
\langle n_{N,Z}\rangle_{gc} & = & e^{\beta\mu_{n}N+\beta\mu_{p}Z}\omega_{N,Z}
\end{eqnarray}
 The chemical potentials $\mu_{n}$ and $\mu_{p}$ are determined by solving two equations $N_{0}=\sum Ne^{\beta\mu_{n}N+\beta\mu_{p}Z}\omega_{N,Z}$ and $Z_{0}=\sum Ze^{\beta\mu_{n}N+\beta\mu_{p}Z}\omega_{N,Z}$. This
amounts to solving for an infinite system but we emphasize that this infinite system can break up into only certain kinds of species as are included in the above two equations. We can look upon the sum on $N$ and $Z$ as a sum over $A$ ($=N+Z$) and a sum over $Z$. In principle $A$ goes from 1 to $\infty$ and for a given $A$, $Z$ can go from 0 to $A$. Here for a given $A$ we restrict $Z$ by the same drip lines used for canonical model.\\
\indent
In both the models, the partition function of a composite having $N$ neutrons and $Z$ protons is a product of two parts: one is due to the the translational motion and the other is the intrinsic partition function of the composite:
\begin{eqnarray}
\omega_{N,Z}=\frac{V}{h^{3}}(2\pi mT)^{3/2}A^{3/2}\times z_{N,Z}(int)
\end{eqnarray}
where $V$ is the volume available for translational motion. Note that $V$ will be less than $V_{f}$, the volume to which the system has expanded at break up (freeze-out volume). We use $V=V_{f}-V_{0}$ , where $V_{0}$ is the normal volume of nucleus with $Z_{0}$ protons and $N_{0}$ neutrons. In this work the temperature and freeze-out volume are kept constant at 5 MeV and $3V_0$ respectively.\\
\indent
We list now the properties of the composites used in this work.  The proton and the neutron are fundamental building blocks thus $z_{1,0}(int)=z_{0,1}(int)=2$ where 2 takes care of the spin degeneracy.  For deuteron, triton, $^3$He and $^4$He we use $z_{N,Z}(int)=(2s_{N,Z}+1)\exp(-\beta E_{N,Z}(gr))$ where $\beta=1/T, E_{N,Z}(gr)$ is the ground state energy of the composite and $(2s_{N,Z}+1)$ is the experimental spin degeneracy of the ground state.  Excited states for these very low mass nuclei are not included.
For mass number $A\ge5$ we use the liquid-drop formula.  For nuclei in isolation, this reads\\
\begin{eqnarray}
z_{N,Z}(int)
&=&\exp\frac{1}{T}[W_0A-\sigma(T)A^{2/3}-a^{*}_c\frac{Z^2}{A^{1/3}}\nonumber\\
&&-C_{sym}\frac{(N-Z)^2}{A}+\frac{T^2A}{\epsilon_0}]
\end{eqnarray}
The expression includes the volume energy [$W_0=15.8$ MeV], the temperature dependent surface energy
[$\sigma(T)=\sigma_{0}\{(T_{c}^2-T^2)/(T_{c}^2+T^2)\}^{5/4}$ with $\sigma_{0}=18.0$ MeV and $T_{c}=18.0$ MeV], the Coulomb energy with Wigner-Seitz approximation [$a^{*}_c=a_{c}\{1-(V_{0}/V_{f})^{1/3}\}$ with $a_{c}=0.72$ MeV] and the symmetry energy ($C_{sym}=23.5$ MeV).  The term $\frac{T^2A}{\epsilon_0}$ ($\epsilon_{0}=16.0$ MeV) represents contribution from excited states since the composites are at a non-zero temperature.\\
\indent
In canonical ensemble though chemical potential does not come into picture directly, but we can define them as $\mu=-\frac{\partial F_t}{\partial N}$. So in our case,
\begin{equation}
\mu_p= F_t(N_0,Z_0-1)-F_t(N_0,Z_0)
\end{equation}
\begin{equation}
\mu_n= F_t(N_0-1,Z_0)-F_t(N_0,Z_0)
\end{equation}
where $F_t(N_0,Z_0)=-TlnQ_{N_0,Z_0}$ is the total free energy.\\
\section{Theoretical framework of isoscaling and isobaric yield ratio method}
Substituting $\omega_{N,Z}$ from Eq. (4) and Eq. (5) in Eq. (3), the average multiplicity from grand canonical ensemble can be written as,
\begin{eqnarray}
\langle n_{N,Z}\rangle_{gc}=\frac{V}{h^3}(2\pi mT)^{3/2}A^{3/2}\nonumber\\
\times\exp[-\frac {{F(N,Z)-{\mu_n}N-{\mu_z}Z}}{T}]
\end{eqnarray}
where $F(N,Z)=-Tlnz_{N,Z}(int)$, whereas $z_{N,Z}(int)$ is given by Eq. 5.
In isobaric yield ratio method, ratio of yields of two different types of fragments having same mass number $A$ but
 different isospin asymmetry $I=N-Z$ and $I^{'}=N^{'}-Z^{'}$ originating from same source \cite{Tsang2} is given by,
\begin{eqnarray}
R[I,I^{'},A]&=& \langle n_{I,A}\rangle_{gc}/\langle n_{I^{'},A}\rangle_{gc}
\end{eqnarray}
With the choice of $I=1$ and $I^{'}=-1$ the ratio will be
\begin{equation}
ln R[1,-1,A]=\frac{\mu_{n}-\mu_{p}}{T}+\frac{a^{*}_c}{T}A^{2/3}
\end{equation}
And isoscaling is the ratio of yields of the same type of fragment $(N,Z)$ originating from two sources having different
 mass number $A_1$ and $A_2$ ($A_2>A_1$
) but same charge $Z_1=Z_2=Z$. From (6),
\begin{eqnarray}
R_{21}&=& \langle {n_2}_{N,Z}\rangle_{gc}/\langle {n_{1}}_{N,Z}\rangle_{gc}\nonumber\\
&=& C\exp(\frac{\mu_{n_2}-\mu_{n_1}}{T}N+\frac{\mu_{z_2}-\mu_{z_1}}{T}Z)\nonumber\\
&=& C\exp(\alpha N+\beta Z)
\end{eqnarray}
In canonical and grand canonical model calculations, $R_{21}$ is calculated from the ratio of
 $\langle {n_2}_{N,Z}\rangle$ and $\langle {n_1}_{N,Z}\rangle$ and then isoscaling parameters $\alpha$ and $\beta$ are obtained by linear fitting of $lnR_{21}$ with $N$ (at constant $Z$) and $Z$ (at constant $N$) respectively.

Since  the above formalisms are valid for equilibrium condition \cite{Mallik5} and secondary decay
affects the equilibrium scenario \cite{Mallik1, Mallik5}, hence in the entire theoretical calculation
 secondary decay is not included.\\
\section{Results and Discussion}
We have studied the dependence of neutron and proton chemical potentials on both source size and source asymmetry and the results are displayed in Fig. 1. In Fig. 1(a), the variation of both $\mu_n$ and $ \mu_p$ with source proton number $Z_0$  is shown for two values of  the asymmetry parameter $y=(N_0-Z_0)/(N_0+Z_0)=$ 0.11 and 0.27. The source size $A_0$ is varied from 44 to 264 ($Z_0$ from 16 to 96) and it is seen that both the neutron and the proton chemical potential remains almost constant as one increases the source size irrespective of the value of $y$. In the same figure results are shown from both the canonical and the grand canonical models and it is seen that the chemical potentials calculated from both the ensembles are almost equal except for the very small sources where they are slightly different. In Fig. 1(b), the variation of chemical potentials with the asymmetry parameter $y$ is shown where $y$ is varied from 0 to 0.33. The source size $A_0$ is kept fixed at 60 whereas $Z_0$ varies  from 20 ($y$=0.33) to 30 ($y$ =0). The change of $\mu_n$ and $\mu_p$ with $y$ is almost linear for both the ensembles and also the results from the canonical and the grand canonical ensembles are more or less  the same except for higher $y$ values where they are slightly different.\\
\begin{figure}[h]
\includegraphics[width=2.8in,height=2.1in,clip]{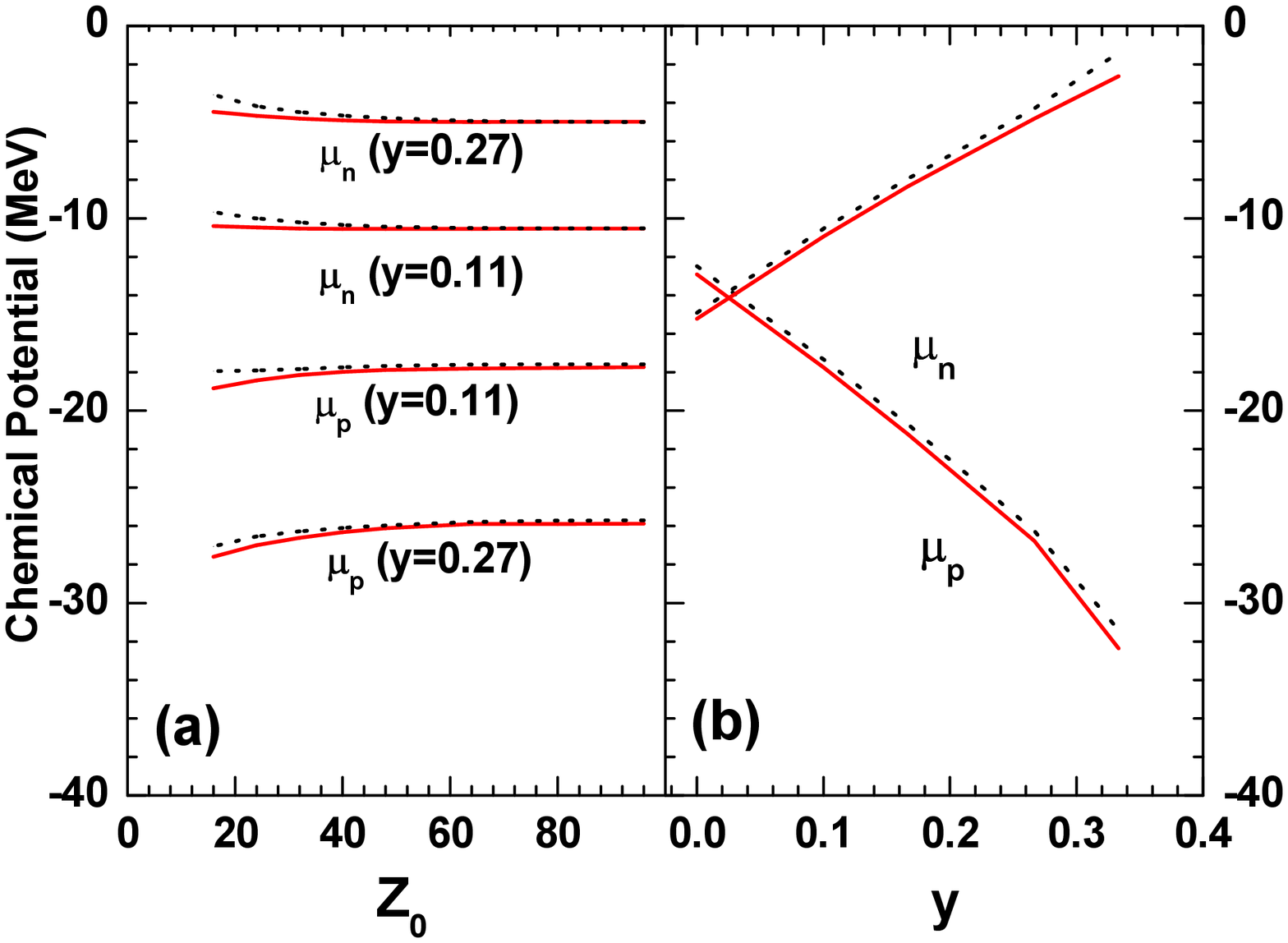}
\caption{(Color online) Variation of proton chemical potential ($\mu_p$) and neutron chemical potential ($\mu_n$) with (a) source size (at constant source asymmetry $y=0.27$ and $0.11$) (b) source asymmetry (for fixed source size $A_0=60$) from canonical (red solid lines) and grand canonical models (black dotted lines).}
\label{fig1}
\end{figure}
\indent
Since it is seen from Fig. 1(a) \& 1(b) that the chemical potentials are almost same for both models for the entire range of source size and source asymmetry, hence for all other results to be presented for these systems,  the chemical potentials obtained from grandcanonical model are being used.\\
\begin{figure}[b]
\includegraphics[width=2.8in,height=3.4in,clip]{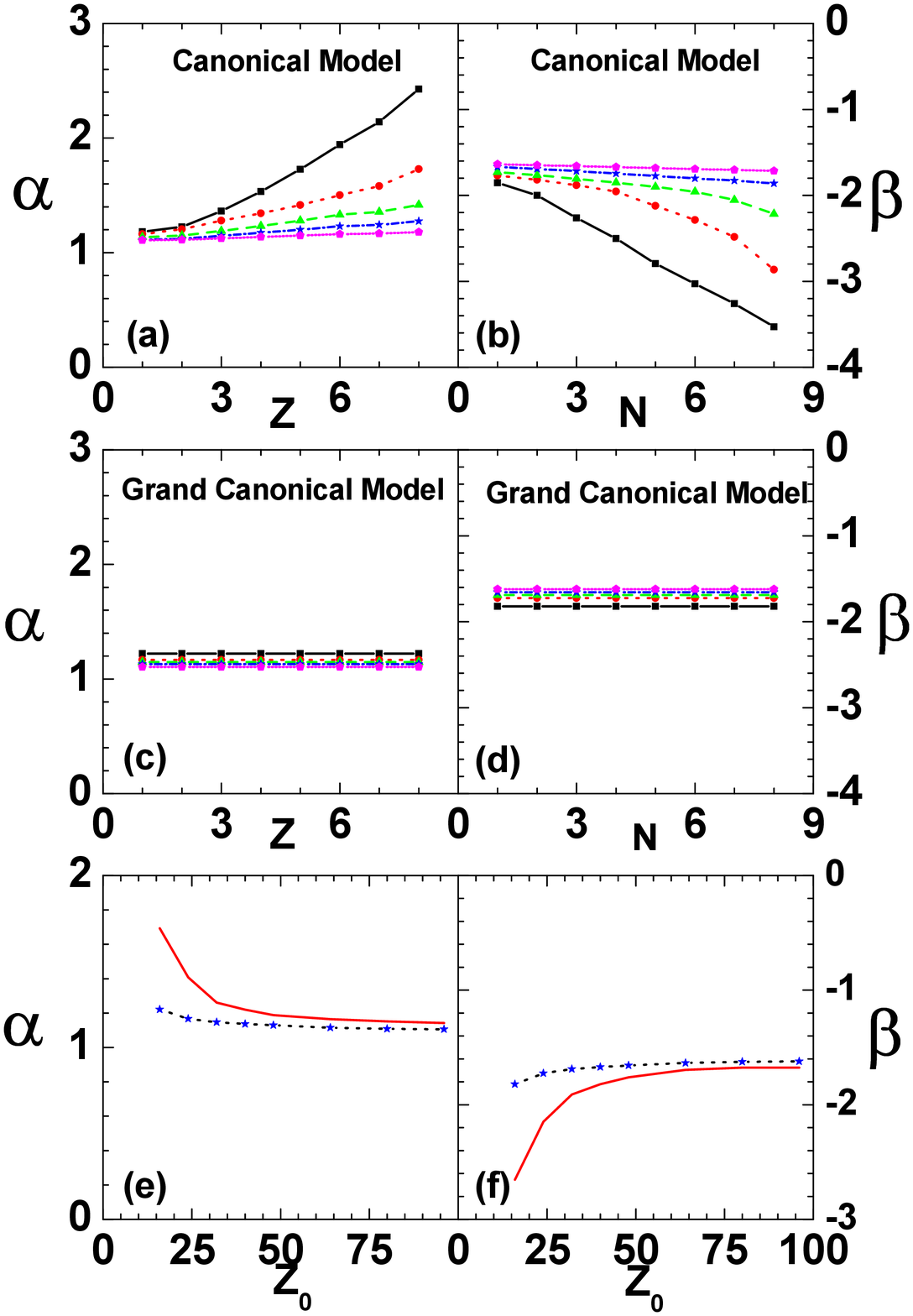}
\caption{(Color online) Variation of isoscaling parameter $\alpha$ ($\beta$)
with fragment proton number $Z$ ($N$) for each pair of sources having same isospin asymmetry
$0.27$ and $0.11$ but different charges $Z_0=16$ (black squares joined by solid lines),
 24 (red circles joined by dotted lines), 32 (green circles joined by dashed lines)
 48 (blue stars joined by dash dotted lines) and 96 (magenta pentagons joined by short dotted lines)
 from canonical [2(a), 2(b)] and grand canonical [2(c), 2(d)] model. 2(e) and 2(f) shows the variation
of the isoscaling parameters $\alpha$ and $\beta$ respectively with source charge ($Z_0$) obtained
from canonical model (red solid lines),
 grand canonical model (black dotted lines) and that calculated
from the formulae $\alpha=(\mu_{n_2}-\mu_{n_1})/T$ and $\beta=(\mu_{p_2}-\mu_{p_1})/T$ (blue stars).}
\label{fig2}
\end{figure}
\indent
The isoscaling parameters $\alpha$ and $\beta$ (Eq. 11) depend only on the difference in
 chemical potentials of  more neutronrich and less neutron rich fragmenting systems and
 on the temperature at freeze-out. For isoscaling studies a pair of sources with same
 $Z_0$ value is required. We have kept the asymmetry value of the more neutron-rich source to
 be 0.27 and that of the less neutron-rich one to be 0.11. The study was done for different
 source sizes ranging from $Z_0$ =16 to $Z_0$ =96, the y values being the same for each pair.
 Since for a particular pair of reaction the difference in chemical potentials are the same,
 therefore it is expected that the isoscaling parameters  $\alpha$ ($\beta$) would remain constant
throughout the entire $Z$ ($N$)  regime of the fragments. It can also be concluded from Fig. 1(a)
that  at constant temperature and same $y$ value of the fragmenting sources, $\mu_p$ and $\mu_n$
are almost independent of the source size. On the contrary, from theoretical calculation by the canonical
 model it is observed  from Fig 2(a) \& 2(b) that $\alpha$ increases with the increase of  fragment proton
 number $Z$ and $\beta$ decreases with increase of fragment neutron number $N$ and the change is more for
 smaller fragmenting sources. On the other hand, $\alpha$ and $\beta$ values calculated from the
 grand canonical model (Fig 2(c) \& 2(d)) are independent of the fragment size. The dependence on
source size is also very small as compared to the canonical results (Fig 2(a) \& 2(b)).Therefore
if we take the average of $\alpha$ (or $\beta$) in the range  of $Z=1$ to $8$ (or $N=1$ to $8$) for
 each source  and compare those with that calculated from the formula $\alpha=(\mu_{n_2}-\mu_{n_1})/T$ or $\beta=(\mu_{p_2}-\mu_{p_1})/T$, then it is observed that the average values obtained from canonical and grand canonical model are different for the smaller fragmenting sources and the difference decreases substantially
as one increases the source size as seen in Fig 2(e) \& 2(f)). The values of the isoscaling parameter
 $\alpha$ and $\beta$  calculated from the slopes of the ratio $R_{21}$ of the grand canonical model coincides exactly with
those calculated from the formula. This is seen from the dotted lines and the stars in Fig 2(e) \& 2(f). This is what is expected since the formulae connecting the isoscaling parameters with the difference in chemical potentials is deduced from the grand canonical ensemble and hence results from the later exactly coincide with those calculated from the formulae.\\
\begin{figure}[t]
\includegraphics[width=2.8in,height=2.8in,clip]{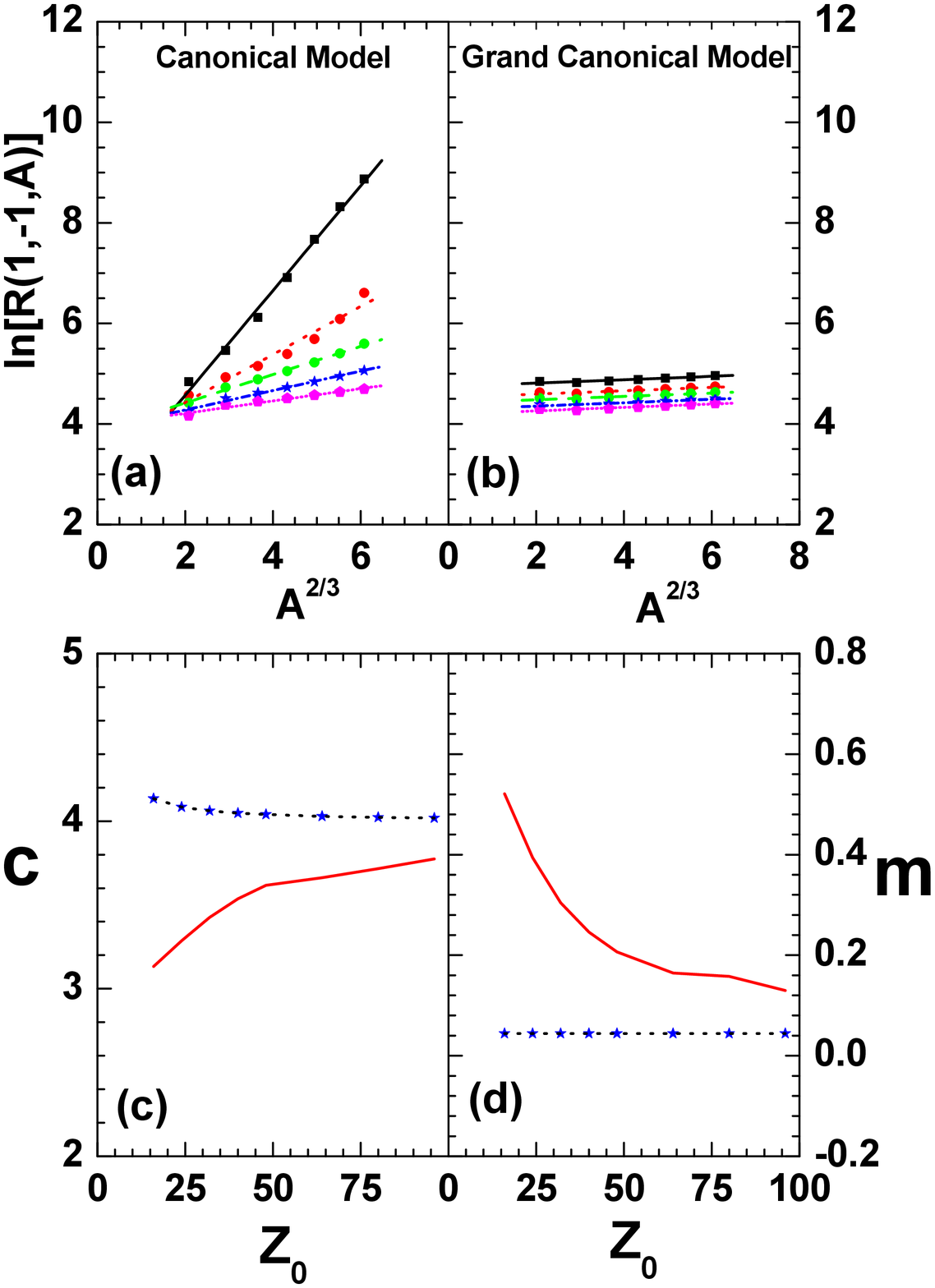}
\caption{(Color online) Variation of isobaric yield ratio $ln R[1,-1,A]$ with $A^{2/3}$ for sources having same isospin asymmetry 0.27 but different charge $Z_0=16$ (black squares), 24 (red circles), 32 (green circles) 48 (blue stars) and 96 (magenta pentagons) from canonical (a) and grand canonical (b) model. Here the lines connecting the points represent the linear fittings. (c) and (d) shows variation of the isobaric yield ratio parameters $c$ and $m$ respectively with source charge ($Z_0$) obtained from canonical model (red solid lines), grand canonical model (black dotted lines) and that calculated from the formulae $c=\Delta\mu/T$ and $m=a^*_c/T$ (blue stars).}
\label{fig3}
\end{figure}
\begin{figure}[t]
\includegraphics[width=2.8in,height=2.8in,clip]{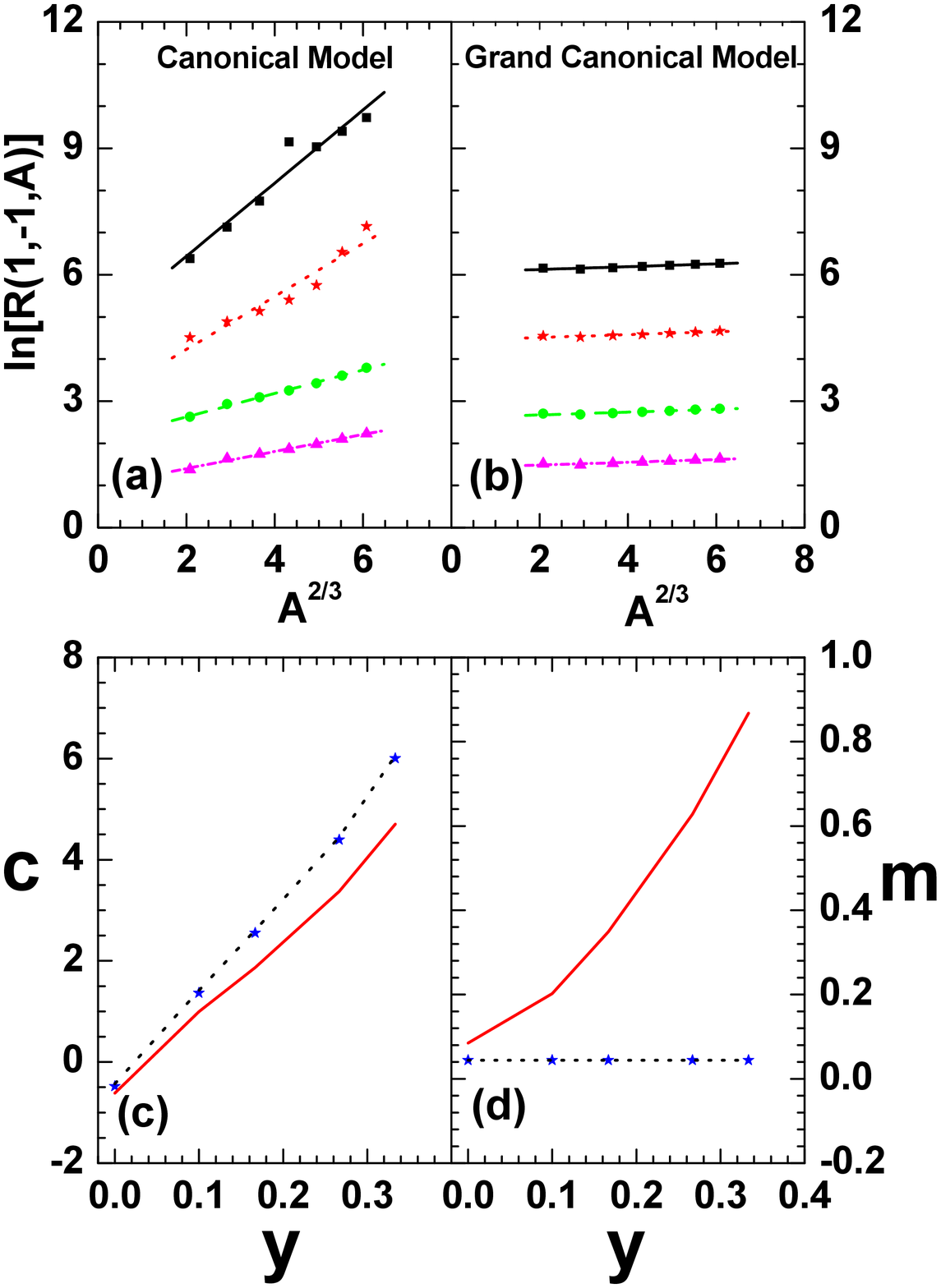}
\caption{(Color online) Variation of isobaric yield ratio $ln R[1,-1,A]$ with $A^{2/3}$ for sources having same mass $A_0=60$ but different isospin asymmetry 0.1 (black squares), 0.17 (red stars), 0.27 (green circles) and 0.33 (magenta triangles) from canonical (a) and grand canonical (b) model. Here the lines connecting the points represent the linear fittings. (c) and (d) shows the variation of isobaric yield ratio parameters $c$ and $m$ respectively with source isospin asymmetry ($y$) obtained from canonical model (red solid lines), grand canonical model (black dotted lines) and that calculated from the formulae $c=\Delta\mu/T$ and $m=a^*_c/T$ (blue stars).}
\label{fig4}
\end{figure}
\indent
In Fig. 3(a) \& 3(b) we show the variation of the isobaric yield ratio $lnR[1,-1,A]$ with the fragment
 sizes where each line represents  a particular source having different size but same isospin asymmetry
 0.27. In the Isobaric yield ratio method, it is observed from Eq. 10 that the quantity $lnR[1,-1,A]$
 calculated for odd  $A$  nuclei (since $N - Z = 1 or -1$), varies linearly with $A^{2/3}$ by
an equation like $y=mx+c$ with $m=a^{*}_c/T$ \&  $c=\Delta\mu/T$
(where $a^{*}_c=a_{c}\{1-(V_{0}/V_{f})^{1/3}\}$ and $\Delta\mu= \mu_n-\mu_p$).
 The difference $\mu_p - \mu_n$  is almost independent of the source size (Fig. 1(a));
 hence c's should be almost equal and m's are exactly equal for all the fragmenting sources.
 Therefore the plot of the ratio $lnR[1,-1,A]$  originating from all the fragmenting sources
 (as used in Fig. 1(a)) should almost coincide. But from the canonical model,
 the variation of $lnR[1,-1,A]$ with $A^{2/3}$ is different for different sources as shown in Fig. 3(a).
 The slopes of the lines from different sources vary with the source size although the slope
should be exactly equal according to the formula. This deviation arises because from Eq. 10
the slope $m=a^*_c/T$ is derived from the grand canonical model and  the same may not hold
 true for the canonical results. The results from the grand canonical model are shown in Fig. 3(b).
Here it is seen that the slopes are exactly equal irrespective of the source size.
The calculated values of the slope $m$ and  the y-intercept $c$ obtained from linear
 fitting of the lines from canonical models (Fig. 3(a)) and those calculated from formula
 $c=\Delta\mu/T$ and $m=a^*_c/T$ are not same for smaller fragmenting sources,
but are close for  the larger sources. This is shown in Fig. 3(c)  \&  3(d).
  The reason for this deviation is that the formulae are derived using
 the grand canonical ensemble and hence they are in general not true for the canonical model results.
 For larger sources the fragmentation is more, therefore the particle number fluctuation
 in grand canonical model is very less \cite{Das1}. In canonical model, particle number is strictly conserved and
there is no such fluctuation. Hence the isoscaling parameters and isobaric yield ratios obtained from canonical and grand canonical model become closer compared to that from the smaller fragmenting sources. The values of $c$ and $m$ obtained by fitting the lines from the grand canonical model (Fig. 3(b)) coincide exactly with the
 values given by the formula. This is shown by the dashed line and the symbols in Fig. 3(c) and Fig. 3(d).\\
\indent
In Fig. 4 we show the effect of variation of the source asymmetry $y$ on the isobaric yield ratio parameters.
In Fig. 4(a) we plot the ratio $lnR[1,-1,A]$ with $A^{2/3}$ where $A$ is the mass number of the fragment.
The different lines on the plot corresponds to different sources with $y$ values ranging from 0.33 to 0.
 The slopes of these lines according to Eq. 10 is equal to $m=a^*_c/T$ and hence should not depend on
its $y$ value. The value of the y-intercept $c$ of these lines will be different since it is
 equal to $(\mu_n-\mu_p)/T$ which depend on $y$ as seen from Fig 1(b). It is seen from Fig. 4(a)
 that the slopes are different for different sources from the canonical model calculation,
 the deviation being more for the source which is more asymmetric. For the grand canonical model(Fig. 4(b)),
 the slopes are exactly equal for each source as expected from the formulae. The values of the parameters
 $c$ and $m$ are plotted in Fig. 4(c) \& 4(d) respectively. It is seen that results from the canonical model
 differs from that of the formulae for higher values of $y$ and they become close as
 $y$ value approaches 0 or in other words the source becomes symmetric. The results from the
 grand canonical ensemble coincide exactly with that from the formulae. It has been already studied
 that results from the canonical and grand canonical models converge more as the fragmenting system
 becomes more symmetric as the particle fluctuation in grand canonical model becomes less in such cases.
 Similar effect is obtained earlier for mass distribution \cite{Mallik4}.\\
\section{Summary and Conclusion}
Isobaric yield ratio method as well as the isoscaling equations connect the liquis drop model parameters
to the ratio of fragment yields from break-up of hot nuclei. These relations are deduced using the grand canonical ensemble
for calculating the fragment yields. Hence results from this ensemble agree exactly with those
 obtained from the equations of isobaric yield rati and isoscaling. On the contrary, results from the canonical ensemble
(which is more suitable for describing finite nculei with conserved mass and charge) deviate from these equations and this deviation
is more pronounced for smaller source size and more asymmtetric nuclei. This is very much expected  since results from both the ensembles differ in
general for finite nculei and are found to converge when fragmentation of the nucleus is more and
it happens for the sources having larger mass or less asymmetry\cite{Mallik4}.\\

\end{document}